\documentclass[prl,superscriptaddress,twocolumn,floatfix,showpacs,letterpaper]{revtex4}
\usepackage{amsmath,amsthm,amssymb,amscd}
\usepackage{mathtools}
\DeclarePairedDelimiter\ceil{\lceil}{\rceil}
\DeclarePairedDelimiter\floor{\lfloor}{\rfloor}

\usepackage{graphicx}
\usepackage{times}
\usepackage{hyperref}
\usepackage{float}
\usepackage{array, multirow}
\usepackage{color}
\usepackage{caption}
\usepackage{booktabs}
\usepackage{enumitem}

\newcommand{\fmtm}{\textit{Fm}$\bar{3}$\textit{m}}
\newcommand{\cmca}{{\em Cmca} }
\newcommand{\immm}{{\em Immm} }
\newcommand{\bu}{\mathbf{u}}
\newcommand{\bv}{\mathbf{v}}
\newcommand{\bw}{\mathbf{w}}
\newcommand{\bt}{\mathbf{t}}
\newcommand{\norm}[1]{\left\lVert #1 \right\rVert}

\begin{document}

\title{Prediction of Chlorine and Fluorine Crystal Structures at High Pressure Using Symmetry Driven Structure Search with Geometric Constraints}

\author{Mark A. Olson} 
\affiliation{Corresponding author: contact at maolson97@berkeley.edu}
\affiliation{Department of Mathematics, University of California, Berkeley, CA 94720}
\author{Shefali Bhatia}
\affiliation{Department of Electrical Engineering and Computer Science, University of California, Berkeley, CA 94720}
\author{Paul Larson}
\affiliation{Department of Mathematics, Miami University, Oxford, OH 45056}
\author{Burkhard Militzer} 
\affiliation{Department of Earth and Planetary Science, University of California, Berkeley, CA 94720}
\affiliation{Department of Astronomy, University of California, Berkeley, CA 94720}

\begin{abstract}
    \begin{center}
        {Abstract}
    \end{center}
    The high-pressure properties of fluorine and chlorine are not yet well understood because both are highly reactive and volatile elements, which has made conducting diamond anvil cell and x-ray diffraction experiments a challenge. Here we use {\it ab initio} methods to search for stable crystal structures of both elements at megabar pressures. We demonstrate how symmetry and geometric constraints can be combined to efficiently generate crystal structures that are composed of diatomic molecules. Our algorithm extends the symmetry driven structure search method [Phys. Rev. B 98 (2018) 174107] by adding constraints for the bond length and the number of atoms in a molecule, while still maintaining generality. As a method of validation, we have tested our approach for dense hydrogen and reproduced the known molecular structures of {\it Cmca}-12 and {\it Cmca}-4. We apply our algorithm to study chlorine and fluorine in the pressure range from 10--4000 GPa while considering crystal structures with up to 40 atoms per unit cell. We predict chlorine to follow the same series of phase transformations as elemental iodine from {\it Cmca} to {\it Immm} to {\it Fm}$\bar{3}${\it m}, but at substantially higher pressures. We predict fluorine to transition from a {\it C2/c} to an {\it Cmca} structure at 70 GPa, to a novel orthorhombic and metallic structure with {\it P$4_2$/mmc} symmetry at 2500 GPa, and finally into its cubic analogue form with {\it Pm}$\bar{3}${\it n} symmetry at 3000 GPa.
\end{abstract}

\maketitle
\section{Introduction}
High pressure experimentation has made important contributions to the fields of geophysics, mineralogy, and conductivity \cite{mao15, mcmillan02, shimomura78}. It has led to an increased understanding of graphene and 2D layered materials~\cite{balandin08, wang2012}, discoveries of new stoichiometries for common materials~\cite{zhang13}, and new developments with respect to the flow of solid state materials in the Earth's mantle, resulting in an improved quantitative evaluation of geodynamics \cite{mao15}. Similarly, high pressure experimentation has led to the synthesis of new materials, such as synthetic diamonds and cubic boron nitride, both of which are super-hard abrasives \cite{mcmillan02}. One important goal of high-pressure experimentation is the investigation of crystal structures and phase diagrams~\cite{han19}.  Such investigations have resulted in a better understanding of the chemical properties of compounds under high pressures; for example, in 1978, Shinomura et al.~\cite{shimomura78} demonstrated that insights into iodine's crystallographic properties illuminated other properties, including density and conductivity.

One challenge related to experimentation at extreme conditions is working with highly reactive, corrosive, or volatile elements and compounds. For such reasons, chlorine and fluorine are two elements which have been historically challenging to work with \cite{schiferl87}. The crystalline structure of solid chlorine has been studied since 1936 when Keesom and Thomas discovered that, at 185$^\circ$C, chlorine assumed a crystal structure with {\it Cmca} symmetry. This finding was reproduced in 1952 at $-$165$^\circ$C in a vacuum by Collin~\cite{collin52}. Still, high pressure experimental results with chlorine appears to be quite rare. The most recent effort to investigate chlorine at high pressures dates back to 1983 when Johannsen et al.~\cite{johannsen83} studied the Raman spectra of solid chlorine up to pressures of 45 GPa. This pressure was not sufficiently high to observe the phase change to a {\it Immm} structure at 133 GPa that we predict in this article. Diamond anvil cell experiments have also been attempted on fluorine, but obtaining reliable results has been difficult \cite{schiferl87}.  

When reaching high pressures and temperatures is challenging with laboratory experiments or when conducting measurements under such extreme conditions is difficult, {\it ab initio} computer simulations have become the preferred approach. \emph{Ab initio} simulations have been combined with a number of different crystal structure search techniques to identify the most stable structure at high pressures. Evolutionary algorithms~\cite{lyakhov2013}, random structure search techniques \cite{pickard2011}, minima hopping~\cite{goedecker04}, simulated annealing~\cite{woodley08}, and metadynamics~\cite{laio02} have all been designed to increase search efficiency. However, these probabilistic methods do not consider major aspects of the existing data for the materials that they are being employed to study, such as the structural symmetry of certain compounds. \cite{domingos18}.

Symmetries are a key component of studying large clusters \cite{wheeler2007, lv2012, oakley13} and crystal structures \cite{wang2010, pickard2011, wang2012, lyakhov2013}. The addition of symmetry constraints has been shown to improve the algorithmic efficiency for systems with many atoms~\cite{domingos18}. In 2011, Pickard and Needs~\cite{pickard2011} presented an algorithm in which {\it $N_{op}$} symmetry operations are chosen randomly, an atom is placed, and its images are generated according to symmetry. In the case of a mirror plane, half of the atoms are placed randomly; the other half are placed as corresponding mirror images. During the following structural relaxation, it is possible for a more symmetric structure to arise, as the final structure may belong to a supergroup of the original group with {\it $N_{op}$} symmetry operations. 

{In 2010 and 2012, Wang et al.~\cite{wang2010, wang2012} introduced an evolutionary algorithm, CALYPSO, based on a particle swarm optimization method. In CALYPSO, a penalty is introduced to prevent the generation of structures from the same space group repeatedly. In the case of TiO$_2$, with classical potentials, it was directly shown that the use of symmetry constraints in this algorithm allowed for more low energy structures to be generated so that approximately half as many generations were necessary to find the optimal structure~\cite{wang2012}. Similarly, in 2013, Lyakhov et al.~\cite{lyakhov2013} showed that implementing symmetry constraints in their evolutionary algorithm, USPEX, improved efficiency for determining the ground-state structures of MgAl$_2$O$_4$ using classical potentials; USPEX employs symmetry constraints by placing atoms in the most general Wyckoff positions and merging atoms onto higher symmetry positions if they fall within a certain cutoff. This led to the discovery of the ground state for Mg$_{24}$Al$_{16}$Si$_{24}$O$_{96}$, a 160 atom unit cell structure which had not been found previously without symmetry constraints. In 2017 and 2018, similar approaches which rely upon the sampling of space groups and Wyckoff positions were independently developed \cite{avery2017, domingos18}. The XtalOpt algorithm first selects a set of space groups, then creates a list of all possible combinations of Wyckoff positions consistent with the given structure composition for each space group \cite{avery2017}. Since this list can become extremely large, the size of the list is reduced by placing similar Wyckoff positions into groups. In the case of TiO$_2$ with classical potentials and 10 chosen space groups, this work demonstrated that the symmetry constraints increased the probability of generating both low energy structures and high energy structures, with an increased average energy overall.}

Domingos et al.'s symmetry driven structure search (SYDSS) algorithm~\cite{domingos18} samples from all 1,506 Wyckoff positions associated with the 230 space groups without generating a list at all. Since there is no need for size restriction, there is no change in the probability for how often certain combinations of Wyckoff positions are chosen. This method effectively predicts a novel NaCl-H$_2$O structure at 3.4 Mbar as well as novel carbon-oxygen compounds at 20 and 44 Mbar. 

Up to a certain pressure, hydrogen, chlorine, fluorine, and iodine, form molecular crystal structures~\cite{naumov16, tsirelson95, gamba87, shimomura78}, in which the atoms are arranged in diatomic molecules. This suggests that one could generate potentially viable structures with improved efficiency by developing an algorithm that places entire molecules of individual atoms. The known low-pressure molecular structures could serve as initial geometries. 

In existing structure search algorithms that employ symmetry constraints, molecular structures are generated routinely. However, the correct geometries are either sampled by chance or emerge during the structural relaxation. The novel H$_2$O-NaCl structure~\cite{domingos18} is an example of the latter case. The structure has \textit{Pnma} symmetry and the typical geometry of a H$_2$O molecule is well preserved. However, this geometry was only recovered during relaxation when the \textit{Pnma} structure was obtained from, e.g., an initial structure with P2$_1$2$_1$2$_1$ symmetry. Initially the atoms were placed according on the Wyckoff positions in that space group. By observing known molecular geometries, it should be possible to more efficiently predict viable structures. Since molecular geometries may undergo small alterations with increasing pressures, one would want to design a structure search algorithm that accepts some variations for low-pressure geometries. This ensures that no potentially viable structure is excluded.

%%%%%%%%%%%
% METHODS %
%%%%%%%%%%%
\section{Methods}

%We chose to pair our molecular SYDSS algorithm with density functional theory %(DFT) in order to first generate an initial set of structures then %subsequently perform structural relaxation.

Our original SYDSS algorithm~\cite{domingos18} generates symmetric crystal structures by sampling from the 230 space groups and placing atoms at the 1,506 associated Wyckoff positions. The probability of generating symmetric structures is consequently very high, but no structure is in principle excluded; structures with $P_1$ symmetry are generated as well. The challenge that we address in this paper is how to combine symmetry and geometric constraints. It would be preferable to have a general-purpose algorithm that generates symmetric crystal structures entirely composed of, e.g., of H$_2$ or H$_2$O molecules, NH$_3$ tetrahedra or even entire benzene rings (C$_6$H$_6$). As we will illustrate below, it is extremely difficult to construct a general-purpose algorithm that places these molecules given the multitude of the 230 space groups, without excluding any particular configurations. 

Let us consider the simplest case: crystal structures composed of diatomic homonuclear molecules like H$_2$, F$_2$, Cl$_2$, Br$_2$, or I$_2$. Additionally, let us assume that the bond length has been fixed to a specific value like the gas-phase bond distance. (One could consider a range of bond lengths but the sampling challenge remains the same.) Furthermore, we require that any nonbonded atom be at least a certain distance, such as 1.3 bond lengths, away from any other atom in the molecule. The goal of our algorithm is to place atoms so that they simultaneously satisfy the bond length constraint and the symmetry operations of the chosen space group. (We require the individual atoms to be on Wyckoff positions, not just the molecular center of mass. The latter would be appropriate to generate quantum solids of freely rotating molecules like the Pa$_3$ structure of the hydrogen but this is not our goal.) 
Sampling configurations with only geometric constraints is straightforward~\cite{ryckaert76}. For diatomic molecules, one could place a single atom first and then sample the position of its molecular partner from the sphere surrounding it. If periodic boundary conditions are applied, positions from multiple spheres may need to be considered. 

If there is a mirror plane in the system, three cases need to be considered: a) the molecule lies in the plane, b) the molecule straddles it, or c) the molecule is located off the plane and has a mirror image. Sampling configurations from these three cases poses no difficulties. 

If there is an $n$-fold rotation axis in the system, again three cases emerge: a) the molecule lies on the axis, b) it straddles it (unlikely for $n>2$), or c) it lies off the axis, in which case there are $n$ images (The alternative of having a ring with $n$ atoms would violate the distance criteria for the non bonded atoms.)

While handling individual symmetry operations is not difficult, we were not able to construct an algorithm that directly samples the atomic positions for all 230 space group while satisfying our bond length constraints, without excluding any configurations in the process. Some space groups include many symmetry operations. In some cases, the bond length constraint may only be satisfied for specific unit cell parameters. While one may still be able to place molecules by selecting one Wyckoff position after the other, such an approach can be considered impractical since the two partners in a molecule may not necessarily come from the same Wyckoff position. If one wanted to generate such a structure step by step, one would need to keep a list of all dangling bonds, with the goal of satisfying them with atoms to be placed later. Because of this complexity, we designed the following alternative approach that circumvents these issues while retaining generality.

We use our original SYDSS algorithm to place all atoms on Wyckoff positions while keeping a list of all free parameters that were sampled, which may include coordinates, lattice parameters and angles. At this point, no bond length criteria have been considered. To determine which atoms form pairs, we apply the Hungarian algorithm~\cite{kuhn55} to solve a linear sum assignment problem. In cases where this algorithm yields trimers or chains, we discard the entire structure and start the process from the beginning. Once the atoms have been paired, we perform a preliminary structural relaxation using the simplex algorithm~\cite{nedler65} which precedes the later DFT optimization. In this preliminary relaxation, we minimize a penalty function which consists of the sum of the following three terms:
\begin{enumerate}[label=(\alph*)]
\item We add penalization term for molecules that are too close together, using $Q_1=(r_{ij}-r_{\rm n.b.~min})^2$ for all pairs of nonbonded atoms that are closer than $r_{\rm n.b.~min} = 1.3 \times r_{\rm bond}$.
\item We add a penalization term for bonds that are too short within a molecule, using term $Q_2=(r_{ij}-r_{\rm bond~min.})^2$ if $r_{ij}<r_{\rm bond~min.}=0.8 \times r_{\rm bond}$. Similarly, we add a penalization term for bonds that are too long, using $Q_2=(r_{ij}-r_{\rm bond~max.})^2$ if $r_{ij}>r_{\rm bond~max.}=1.2 \times r_{\rm bond}$.
\item We add a term $Q_3 = ((V-V_T)/V_T)^2$ that compares the cell volume, $V$, with its target value, $V_T$. 
\end{enumerate}
Then, only if the preliminary optimization succeeds in reducing the penalty function to a very small value like 10$^{-5}$, we pass the structure on to the DFT optimization, a process which is 10$^5$ times more computationally expensive than our preliminary stage. The goal of the preliminary optimization is to only relax configurations that are already molecular with DFT. The penalty function was chosen so that no configuration that is composed of reasonably well-defined molecules is excluded. Since we only optimize the free parameters on our list, the existing crystal symmetries are preserved. We have thus put forth a general algorithm that satisfies the symmetry and geometric constraints of diatomic molecules.

\begin{figure}[t]
    \includegraphics[width=.5\textwidth]{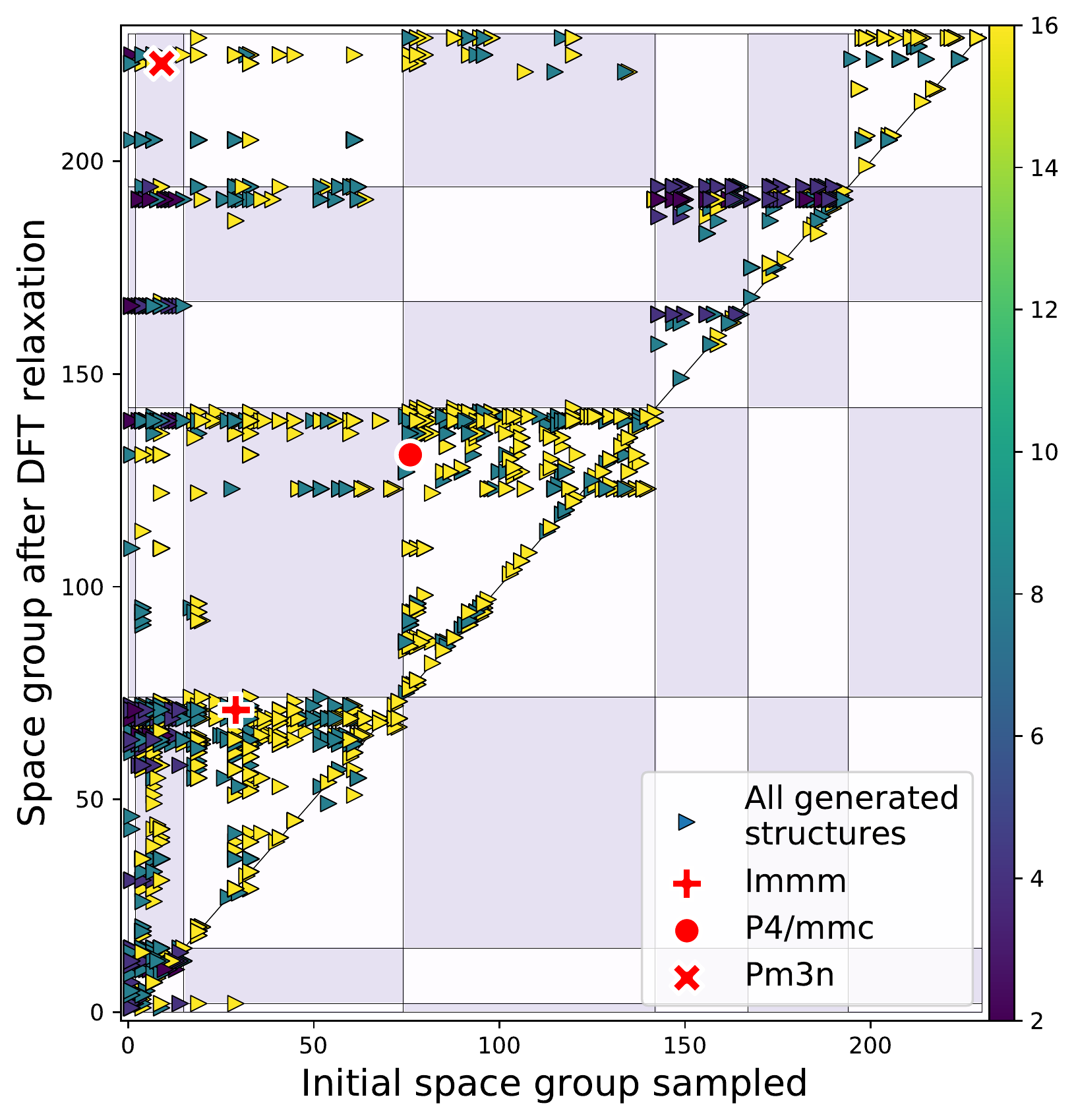}
    \caption{The space group sampled for fluorine by the SYDSS algorithm is plotted against the space group that each structure attained after DFT relaxation. The red symbols show a path for each of the novel structures we found. The colors of the other symbols represent the number of atoms in the unit cell, as specified by the color bar on the right.}
    \label{fig:fig4}
\end{figure}

There are still a small number of parameters that need to be be adjusted manually so that the resulting structures are well suited for DFT calculations. These parameters include unit cell volume, acceptable bond length range, acceptable distances for non-bonded atoms, and number of atoms per cell. We chose unit cell volumes at each pressure based on a reference structure which we relax with DFT calculations. For chlorine specifically, we used the $\alpha$-structure~\cite{collin52}. 

We also used widely available information about the bond lengths of H$_2$, F$_2$, and Cl$_2$ in order to optimize the bond lengths in the SYDSS algorithm. For fluorine, for example, this distance was 1.42 \AA ~\cite{libre19}. 

Using these parameters, we generated a large number of candidate structures for each element, containing between two to forty atoms per cell in order to include a wide range of potential complexity. Only even numbers of atoms were considered because we were interested in structures composed of diatomic molecules. Throughout the process of adjusting these parameters, we frequently checked the variety of the structures that the SYDSS algorithm was generating in order to ensure that we were not systematically excluding any structure classes.

All of the structures that we generated were relaxed with DFT calculations using the Vienna Ab Initio Simulation Package (VASP)~\cite{kresse98}. We used Perdew–Burke-Ernzerhof (PBE) functionals with the projector augmented wave (PAW) method \cite{perdew96}. A conservative value of 1100 eV was chosen for the plane wave energy cutoff. The Brillioun zone was sampled with Monkhorst-Pack $k$-point grids that were adapted to the differing cell dimensions with the VASP k-spacing parameter, $s_k$. The number of $k$-points is obtained by rounding up the ratio $|b_i| / s_k$ to the next integer~\cite{sholl09,vasp19}. $b_i$ is the reciprocal lattice vector. In very heterogeneous unit cells, this ensures that a denser k-point grid is used along the short directions. 

Initial relaxations were performed with a k-spacing parameter of 0.65, followed by a subsequent relaxation with a k-spacing of 0.4. After these two relaxations, a final set of calculations with $s_k=0.25$ was performed on a selection of the structures with the lowest enthalpies, representing structures from 8 different space groups. This final $s_k$ value was chosen based on convergence tests which indicated that a decrease of 0.05 in the k-spacing parameter would result in an enthalpy change of 0.0019\%. 

During DFT relaxation, existing symmetries (as established by the SYDSS algorithm) are preserved, but since atoms may shift to more symmetric positions, the overall symmetry and space group may still change. By comparing the enthalpies of the relaxed structures to each other, we can find the most stable structure of the material at the specified conditions \cite{sholl09}.

%%%%%%%%%%%%%
%% Results %%
%%%%%%%%%%%%%
\section{Results}

\begin{figure}[t!]
    \includegraphics[width=.45\textwidth]{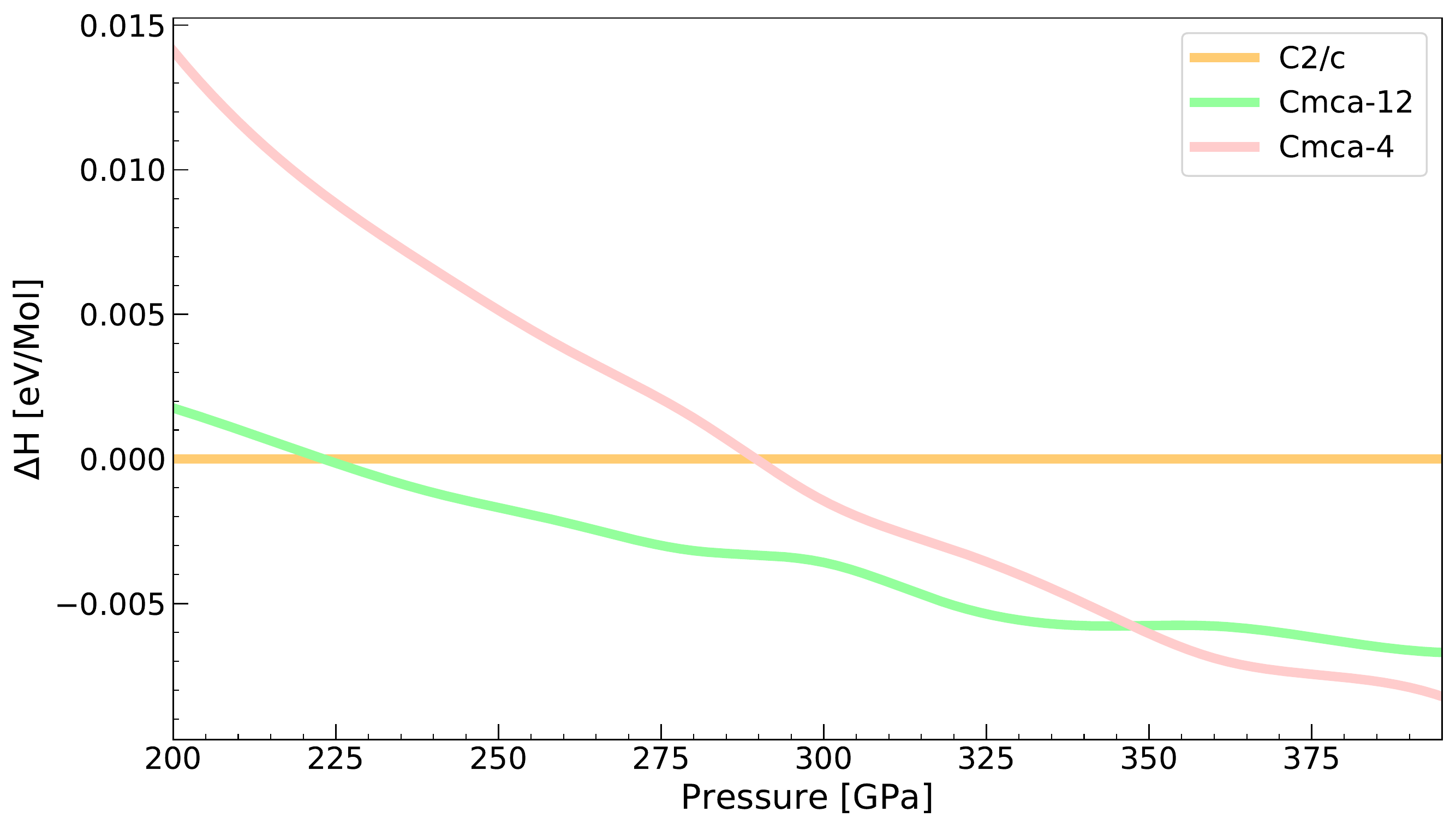}
    \caption{Enthalpy difference of hydrogen with respect to the {\em C2/c} structure. Consistent with results from Refs.~\cite{pickard07,johnson00}, we predict hydrogen to transition to two structures with {\em Cmca} symmetry and 12 and 4 atoms per unit cell at pressures of 225 and 345 GPa, respectively.}
    \label{fig:fig6}
\end{figure}

 Comparing the distribution of the structures generated using the prior version of the SYDSS algorithm - which did not exclude any potential atom configurations - to that of our new version, we can see that the new algorithm results in a less uniform distribution of initial space groups generated. Up to 26\% of generated structures had $P_1$ symmetry - as compared to 10\% in the original algorithm \cite{domingos18}. We did not find, however, that this prevented the potential sampling of structures from all space groups - as was the case for fluorine, shown in Fig \ref{fig:fig4}.
 
 % Chlorine many structure comparison %
\begin{figure}[!t]
    \includegraphics[width=\linewidth]{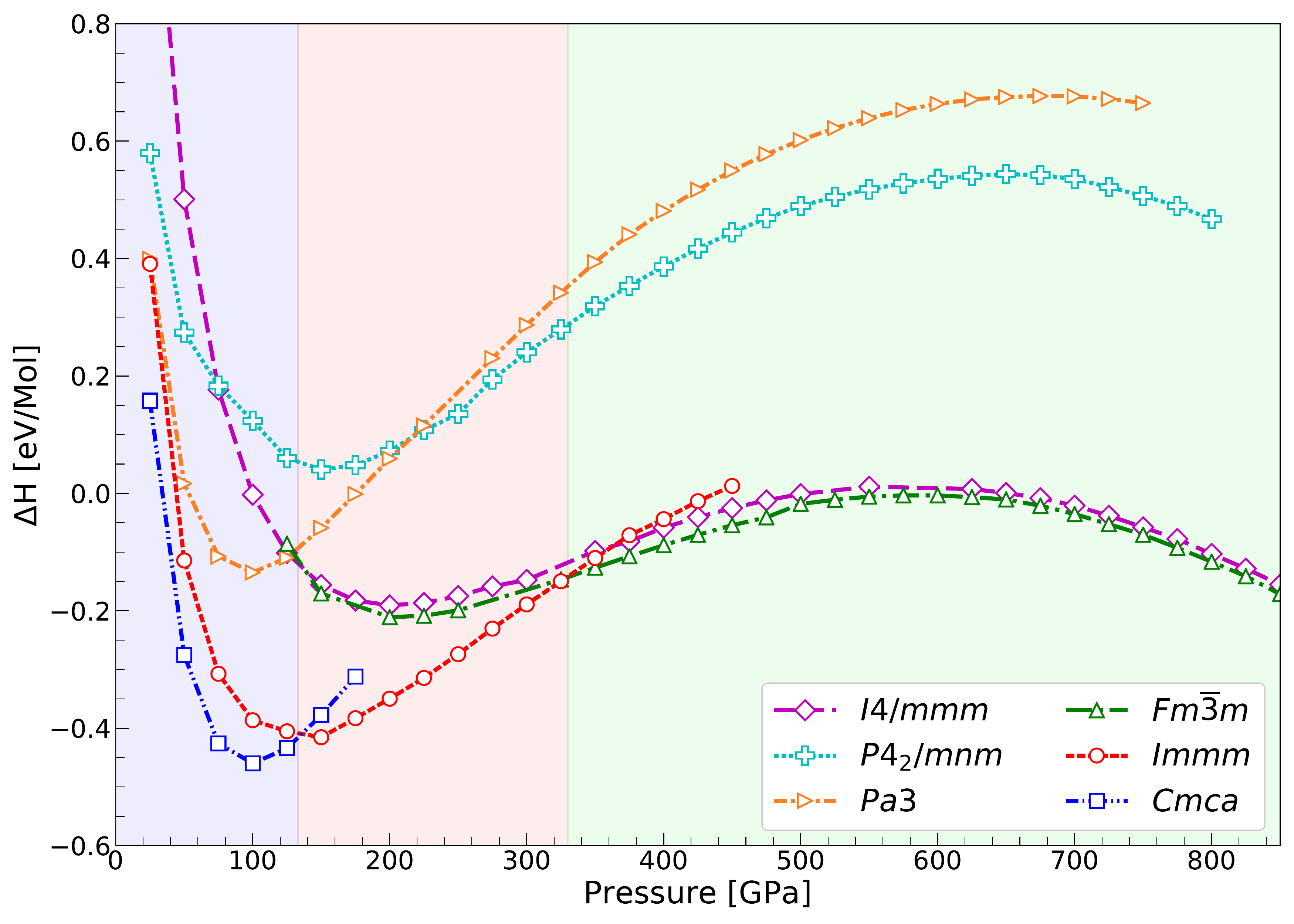}
    \caption{Difference in enthalpy between eight of our energetically competitive chlorine structures and the function $H(P) = 1.07P^{0.65} -6.45$ (in units of GPa and eV/FU) that was obtained by regression. The shaded regions show the pressure intervals where we predict the \cmca, \immm, and \fmtm structures to be stable.}
    \label{fig:fig2}
\end{figure}

In order to test whether our algorithm could generate structures that led to reliable results, we studied dense hydrogen, which is known to form several molecular crystal structures at megabar pressures~\cite{pickard07, johnson00}. Using DFT, we generated and relaxed approximately 65,000 hydrogen structures with 2 to 40 atoms per unit cell within the pressure range of 10 to 1000 GPa. As illustrated in Fig.~\ref{fig:fig6}, we reproduced the findings by Pickard and Needs \cite{pickard07} and Johnson and Ashcroft~\cite{johnson00} - who predicted hydrogen to transition from a structure of $C2/c$ symmetry with 12 atoms per unit cell to a structure of {\em Cmca} symmetry with 12 atoms per unit cell at 225 GPa, and then to another structure of {\em Cmca} symmetry but with 4 atoms per unit cell at 345 GPa. Even though we did not find any new hydrogen structures, our ability to reproduce these results demonstrates that our method is effective in generating physical and energetically competitive molecular structures.

We generated more than 75,000 chlorine structures. After the preliminary molecular optimization was complete, we generated structures in all but two of the 230 space groups ({\em Pmmm} and {\em P222} structures were not generated by the algorithm). This is a more complete set than was generated using the original SYDSSS algorithm on NaCl-H$_2$O, where many space groups were never produced \cite{domingos18}. After DFT relaxations were performed, the distribution of the structures became more highly concentrated around certain space groups. More than 13\% of structural relaxation resulted in structures with {\em I4/mmm} symmetry. A further 8\% were found to relax to structures with {\em Pm$\bar{3}$m} symmetry, 7\% each to structures with {\em P6/mmm} and {\em Im$\bar{3}$m} symmetry, and 6\% with \fmtm ~symmetry. 79 space groups were not represented among the set of relaxed structures. 

Comparing our relaxed structures with the chlorine $\alpha$-structure, we {determine that} two structures emerge at higher pressures than have been investigated with laboratory experiments~\cite{collin52}. At 133 GPa, we {found} the $\alpha$-structure to transition to a highly symmetric body centered orthorhombic structure with {\em Immm} symmetry. The unit cell dimensions are $a = 2.12$ \AA, $b = 2.26$ \AA, and $c = 4.01$ \AA. The atoms are placed on Wyckoff position $a$. This structure is no longer diatomic but is comprised of chains of atoms, where each atom is equidistant to its two neighbors. At 330 GPa, we {found} chlorine to undergo a transition to a face centered cubic (FCC) structure with \fmtm~symmetry. The cell vector is $2.85$ \AA~at this pressure. 

% Chlorine Transition 133 GPa, 330 GPa%
\begin{figure}[!t]
    \includegraphics[width=\linewidth]{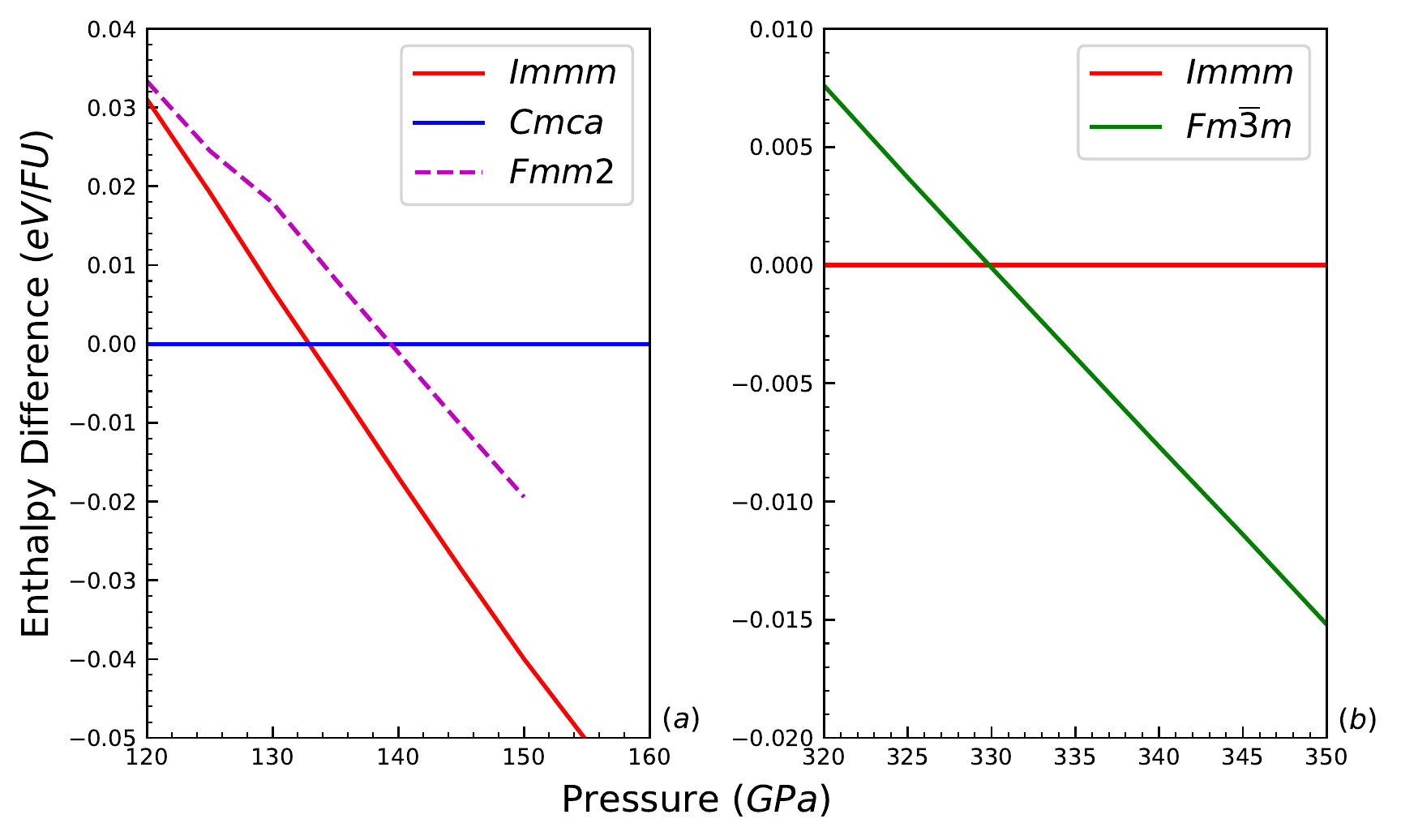}
    \caption{Enthalpy differences between two chlorine structures specified in the caption. At 133 GPa, chlorine transitions from the {\em Cmca} $\alpha$-structure to a body centered orthorhombic structure with {\em Immm} symmetry. At 330 GPa, we predict another transformation to a FCC structures with \fmtm~symmetry. Both transition are accessible with diamond anvil cell experiments. The $Fmm2$ structure from Li et al. \cite{li12} is shown in purple.
    }
    \label{fig:fig3}
\end{figure}

%%%%%%%%%%%%%%%%%%%%%%%%%%%%%%%%%%%%%%%%%%%%
% Visualization of the fluorine structures %
%%%%%%%%%%%%%%%%%%%%%%%%%%%%%%%%%%%%%%%%%%%%
\begin{figure}[t]
    \centering
    \includegraphics[width=2.7in]{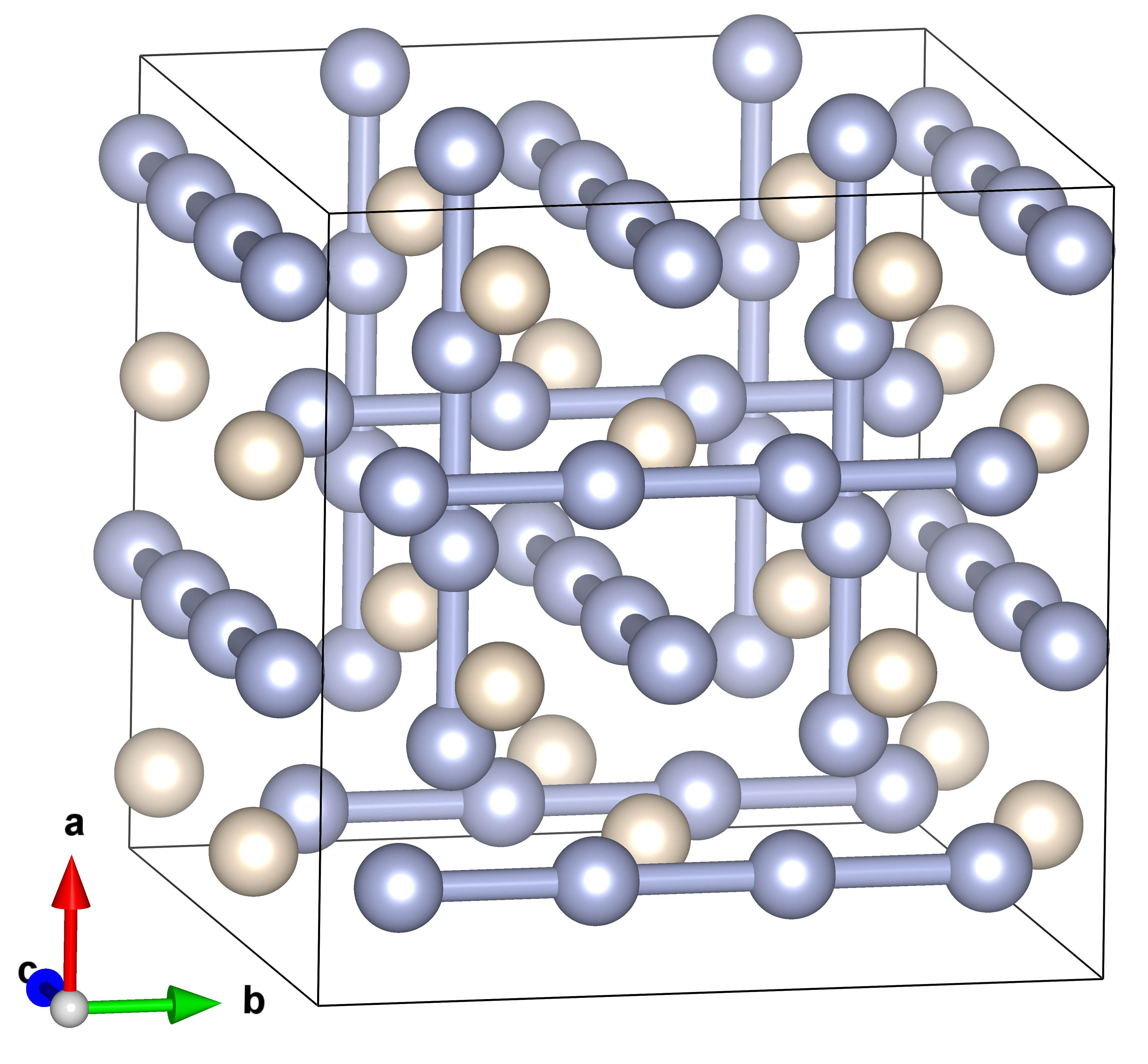} \newline
    (a)
    \includegraphics[width=3.3in]{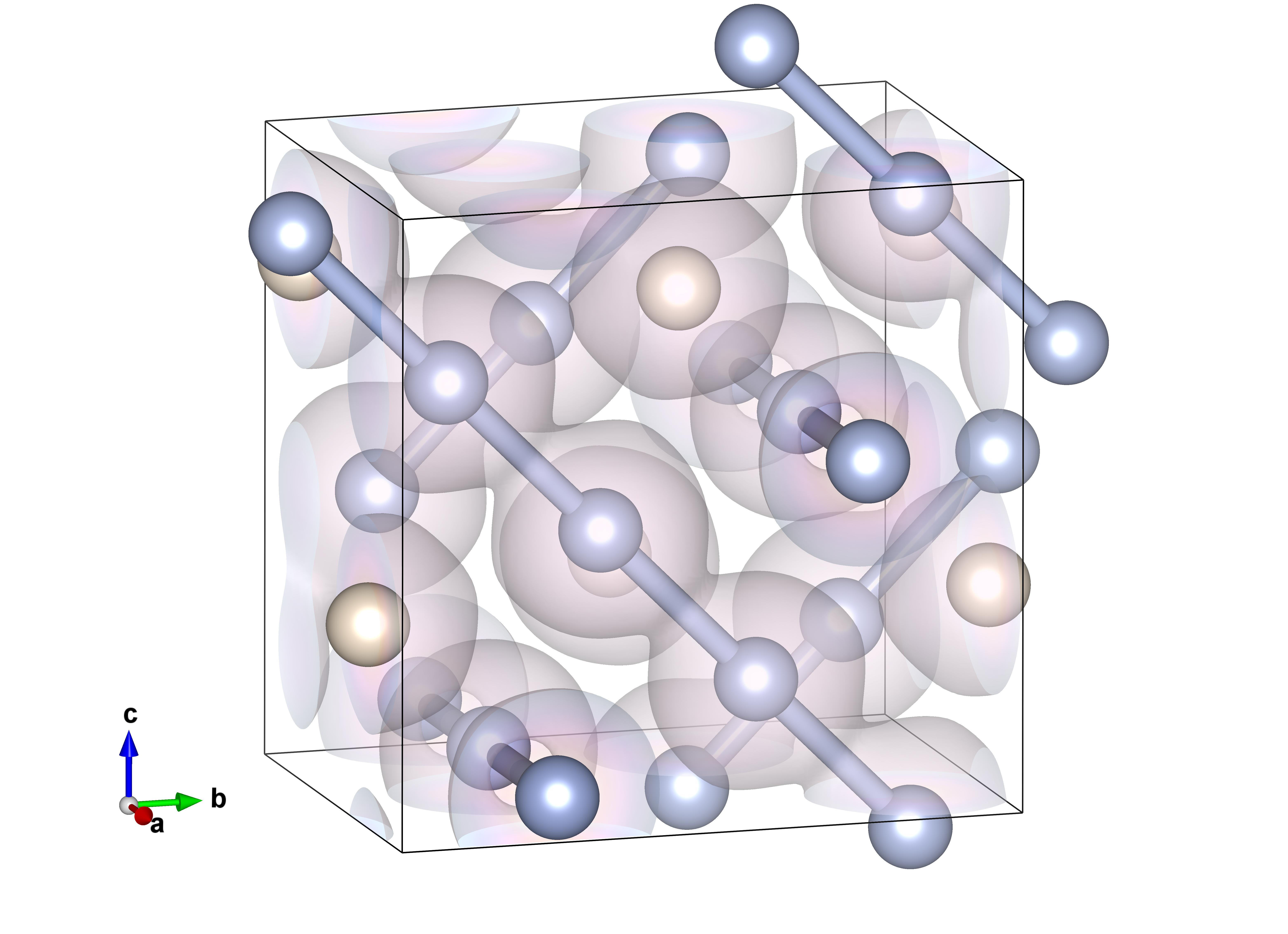}
    (b)
    \label{fig:fig5}
    \caption{The fluorine $Pm\bar{3}n$ structure at 3100 GPa. (b) shows the primitive unit cell while, in (a), a super cell with vectors $a_{\rm ss}=a+b$, $b_{\rm ss}=a-b$, $c_{\rm ss}=2c$ has been constructed to better show the chains that contain 3/4 of the atoms. A lighter color has been used for the remaining atoms reside in the voids in between the chains. In (b), we show surfaces of constant charge density to demonstrate that the chains are separated from each other and that remaining atoms are not bonded.}
\end{figure}

This sequence of transitions is not unexpected because the same sequence was determined for iodine through x-ray diffraction in diamond anvil cells \cite{vonnegut36}. However, we predict these phase changes to occur at much higher pressures for chlorine than for iodine, as our summary in Tab.~\ref{tab:tab1} shows.

% Table of Transition Pressures %
\begin{table}[ht]
    \begin{tabular*}{.95\linewidth}{c@{\extracolsep{\fill}}ccc}
        \hline \hline
        \multirow{2}{*}{Structure} & \multicolumn{3}{c}{Pressure ranges (GPa) for different elements}\\ %\cline{2-4}
                                  & Fluorine      & Chlorine    & Iodine \\ \hline
        $\alpha$ - \textit{C2/c}  & $0-70$        & $\times$    & $\times$ \\
        $\alpha$ - \textit{Cmca}  & $70 - 2500$   & $0-133$     & $ 0-45$ \\
        \textit{Immm}             & $\times$      & $133 - 330$ & $45 - 55$ \\
        FCC - \textit{\fmtm}      & $\times$      & $> 330$     & $> 55$  \\
        $P4_2/mmc$              & $2500 - 3000$ & $\times$    & $\times$\\
        {\it Pm}$\bar{3}${\it n}  & $>3000$       & $\times$    & $\times$\\
        \hline \hline
    \end{tabular*}
    \caption{Pressure ranges where various fluorine, chlorine and iodine crystal structures are predicted to be stable at $T=0$ K. Chlorine and iodine undergo the same sequences of transformations from {\em Cmca} to a body-centered orthorhombic ({\em Immm}) structure, and then to the face-centered cubic (\fmtm) structure but at different pressures. Fluorine behaves differently but still forms the {\em Cmca} structure at intermediate pressures.}
    \label{tab:tab1}
\end{table}

Two energetically competitive structures with {\it P4/mnm} and {\it Pa$\bar{3}$} symmetry were frequently generated. Both are composed of diatomic molecules. As figure~\ref{fig:fig3} shows, they never become energetically competitive at $T=0$ K but it may be worthwhile to investigate whether they become entropy-stabilized at higher temperatures. 

\begin{figure}[t]
    \includegraphics[width=0.99\linewidth]{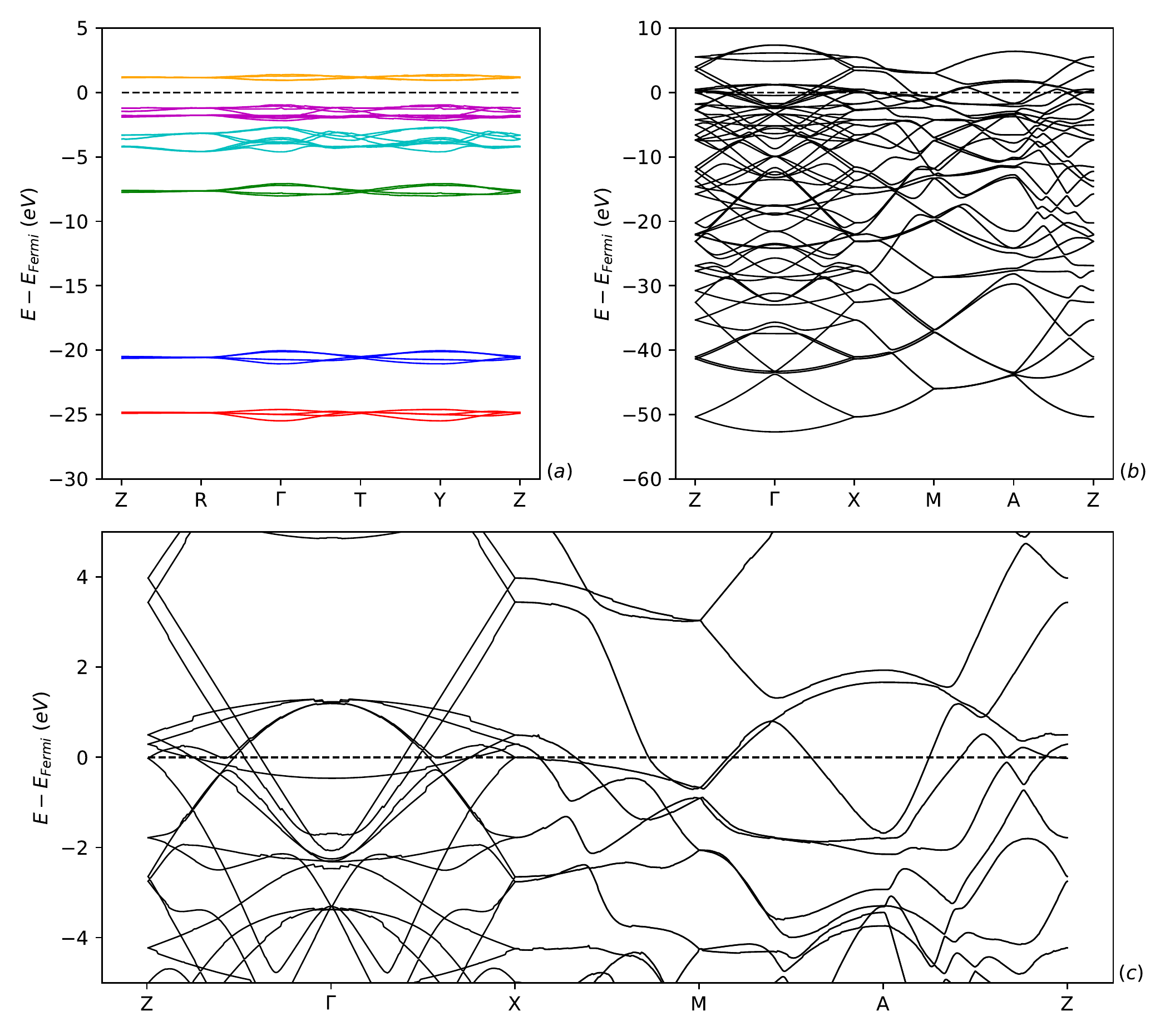}
    \caption{We compare the electronic band structure of two fluorine phases. (a) shows the insulating \cmca structure. The 2s and 2p bands are well separated and system has a band gap of 1.88 eV. The (b) and (c) show the band structure of the metallic $P4_2/mmc$ structure. Its 2s and 2p bands have all hyperdized and the band gap has closed.}
    \label{fig:fig7}
\end{figure}

Our results are in contrast to those of Li et al. \cite{li12}, who predicted the chlorine \cmca structure to transition to a $Fmm2$ structure at 142 GPa, an \immm structure at 157 GPa, and finally to a \fmtm~ structure at 372 GPa. Our results, however, did not indicate that the $Fmm2$ structure was energetically competitive. In addition, the transition pressures that Li et al. presents for \immm and \fmtm ~are notably higher than those we predict. Since we predict the \cmca-to\immm transition to occur already at 133 GPa, we do not find any pressure interval where the $Fm2m$ structure became stable.

We find that fluorine behaves differently from chlorine and iodine. First, we note that the low pressure monoclinic $\alpha$-phase of solid fluorine does not have \cmca but rather $C2/c$ symmetry~\cite{schiferl87}. At 70 GPa, we predict fluorine transitions to the same \cmca structure that we had encountered with chlorine and iodine. After this transition, fluorine does not follow the pattern established by chlorine and iodine. Instead we predict fluorine to remain in the \cmca structure until a pressure of 2500 GPa, at which it transforms to a novel tetragonal structure with $P4_2/mmc$ symmetry that has not been seen for other materials. At 3000 GPa, this structure becomes more symmetric and changes into its cubic analogue form with $Pm\bar{3}n$ symmetry. In both structures, 3/4 of the atoms are arranged in chains that run parallel to the lattice vectors, while the remaining atoms reside in the voids between the chains. In both structures, the atoms occupy the same Wyckoff positions {\em a} and {\em d}. At 2750 GPa, the $P4_2/mmc$ structure has unit vectors $a = 2.69$ \AA, $b = c = 2.60$ \AA. The cubic $Pm\bar{3}n$ structure has a cell vector length of $a = 2.59$\AA~at 3150 GPa. We also find that the $P4_2/mmc$ and $Pm\bar{3}n$ structures are metallic as figure \ref{fig:fig7} shows. In the $Cmca$ structure, the two 2s bands are separated by $\sim15$ eV from and the six 2p bands. The two highest 2p bands are separated by a gap of 1.88 eV. In the $P4_2/mmc$ structure, all 2s and 2p bands have hyperdized and the band gap has disappeared rendering the structure metallic.

This result goes beyond that found by Lv et al. in 2017 \cite{lv17}, where they performed simulations on fluorine at pressures up to 100 GPa, and predicted a curve where fluorine transitions to the \cmca structure at 8 GPa. A key area where our results differ is that we predict that the \cmca transition occurs at much higher pressure, at 70 GPa.

%%%%%%%%%%%%%%
% Conclusion %
%%%%%%%%%%%%%%
\section{CONCLUSION}

We studied the behavior of halogens at megabar pressures with {\it ab initio} computer simulations. We predict chlorine to follow the same sequence of structural transformation as iodine: from an $\alpha$ structure with {\it Cmca} symmetry, to a body-centered orthorhombic structure, and then to a face-centered cubic structure. While these transitions occur at 45 and 55 GPa for iodine, we predict the same transitions to occur at 133 and 330 GPa respectively for chlorine, which renders them accessible to diamond anvil cell experiments. 

Furthermore, we found that fluorine exhibits different behavior from that of iodine or chlorine. We predict fluorine to transform from a $C2/c$ structure at ambient pressure into a C face-centered orthorhombic structure at 70 GPa, and then to two novel metallic structures with $P4_2/mmc$ and $Pm\bar{3}n$ symmetry at 2500 and 3000 GPa, respectively. The chemical bonding in these structures is unusual. Three in four atoms are aligned in an orthogonal set of chains that span all three dimensions, while the remaining atoms are locked into the remaining voids.

We obtained these results using a novel algorithm to generate symmetric crystal structures composed of homonuclear diatomic molecules. A generalization to heterogeneous diatomic molecules is straightforward. The algorithm can also be generalized to sample crystal structures of high symmetry that are composed of triatomic molecules like H$_2$O. After placing the H and O atoms on various Wyckoff positions in the crystal, one would perform a preliminary relation to satisfy O-H and H-H bond length constraints - as we have done for diatomic molecules in this article - before the resulting candidate structures are optimized further with DFT. Our approach may in principle also be applicable to sample configurations that are composed of larger molecules like NH$_3$. However, the efficiency remains be tested because our current algorithm does not enhance to the sampling of configurations where 3 H atoms are placed in the vicinity of every N atom. So, further work will be needed to derive crystal structure prediction algorithms that combine crystal symmetries and arbitrary geometric constrains of large molecules and clusters.

Through comparison to the observed structures in iodine, the novel structures that we have discovered for chlorine and fluorine serve to demonstrate the existence of a relationship regarding the behavior of halogen molecules subjected to extreme pressures. Since we predict that fluorine will undergo further phase changes at very high pressures, it is possible that other halogens may experience similar phase changes as well; this is a research path that may be valuable to continue upon.

\section*{Acknowledgments}

This work was in part supported by the National Science
  Foundation-Department of Energy (DOE) partnership for plasma science
  and engineering (grant DE-SC0016248) and  by the DOE-National Nuclear
  Security Administration (grant DE-NA0003842). Computational support was
  provided the National Energy Research Scientific
  Computing Center.
  PL acknowledges support from the NSF under grant DNS-1764320.
  We thank Rustin Domingos for comments.
  
\section*{Data Availability}
The data of this study are available from the corresponding author upon request.

\section*{Appendix}

In our crystal structures generation algorithm, we frequently need to determine the minimum distance between atom A and atom B or any of its periodic images, which is an instance of a closest vector problem \cite{micciancio2005}. While this is straightforward in nearly cubic cells, it is more challenging in arbitrary unit cell geometries. Before we determine minimum distance between two atoms, we address the simpler problem of finding the closest image of a given atom in periodic boundary conditions. For this derivation we assume the lattice vectors have been scaled and rotated to take the form $\bu = (1,0,0)$, $\bv = (a,b,0)$ and $\bw = (c,d,e)$, with $b$ and $e$ both nonzero. We want to find
integers $i_{*}$, $j_{*}$ and $k_{*}$ (not all $0$) to minimize 
\begin{equation}
    f(i,j,k) = \norm{i\bu + j\bv + k\bw}.
\end{equation}
Since $f(i,j,k) = f(-i, -j, -k)$, we may assume that $k_{*} \geq 0$. So we check each value $k$ from $0$ to $\floor{1/e}$. Since $f(i,j,k) \geq |ke|$ and $f(1,0,0) = 1$, $|k_{*}e| \leq 1$, so $k_{*} \leq \floor{1/e}$. 

Let us write $j_{k}$ for the optimal value of $j$ given $k$. Since $f(i,j,k) \geq |kd + jb|$, for each fixed value of $k$,
$|kd + j_{k}b| \leq 1$, so 
\begin{equation}
\ceil{-(1 + kd)/b} \leq j_{k} \leq \floor{(1 - kd)/b}.
\end{equation}

For given $k$, this defines a range of $j$ values to check. For given values for $k$ and $j$, the optimal value for $i$ is always the negative of the nearest integer to $kc + ja$, which means only this one $i$ value needs to be checked. 

This condition for $i$ combined with the ranges for $k$ and $j_k$ enable us to define an algorithm that finds the optimal triple $i_{*}$, $j_{*}$ and $k_{*}$. 
The total number of triples to check according to this algorithm is
\begin{equation}
\sum_{k=0}^{\floor{1/e}} (\floor{(1 - kd)/b} - \ceil{-(1+kd)/b} + 1).
\end{equation}
It may be that all of these triples give values greater than $1$, in which case $(1,0,0)$ is the optimal triple. 

At the cost of a more complex formula, one could improve the number of $j$ values that need to be checked for each $k$ value, since
$f(i,j,k) \geq \sqrt{(kd + jb)^{2}  + (ke)^{2}}$, so
$|kd + j_{k}b| \leq \sqrt{1 - (ke)^{2}}$.

After solving this simplified problem, we now determine the minimum distance from atom A to atom B or any of its periodic images. We again assume cell vectors $\bu = (1,0,0)$, $\bv = (a,b,0)$ and $\bw = (c,d,e)$, with $b$ and $e$ nonzero, but there is one additional
vector $\bt = (x,y,z)$ that represents the direct difference between atom vectors A and B. We want to find
integers $i_{*}$, $j_{*}$ and $k_{*}$ (possibly all $0$) to minimize 
\begin{equation}
g(i,j,k)  = \norm{i\bu + j\bv + k\bw - \bt}.
\end{equation}
As in the first part, we can find an upper bound for the min by considering any one choice for $i$, $j$ and $k$.
Or we can let $D$ be $\min \{ g(i,j,k) : i,j,k \in \{0,1\}\}$.

Since $g(i,j,k) \geq |ke-z|$, $|k_{*}e - z| \leq D$, so
\begin{equation}
\ceil{(z - D)/e} \leq k_{*} \leq \floor{(z + D)/e}.
\end{equation}
Again, let us write $j_{k}$ for the optimal value of $j$ given $k$.
Since 
\begin{equation}
   g(i,j,k) \geq \sqrt{(jb + kd - y)^{2} + (ke - z)^{2}},
\end{equation}
\begin{equation}
   \sqrt{(j_{k}b + kd - y)^{2} + (ke - z)^{2}} \leq D,
\end{equation}
which defines a range of $j$ values to be checked for a given $k$:
\begin{equation}
    \ceil{(y - kd - l)/b} \leq j_{k} \leq \floor{(y - kd + l)/b}.
\end{equation}
with $l = \sqrt{D^{2} - (ke - z)^{2}}$. 
For each given pair of values $k,j$, the optimal value for $i$ is the negative of the closest integer to $ja + kc - x$. This condition for $i$ combined with the ranges for $k_*$ and $j_k$ again define an algorithm the finds minimum image distance between any images of atoms A and B in arbitrary unit cells with periodic boundary conditions. 
 
We implemented this algorithm and compared the predictions with the empirical algorithm in Ref.~\cite{militzer16} that successively shortens the cell vectors before it compares the distance between atom A and only 3$^3$ images of atom B. For these modified cell vectors, only $i$, $j$, and $k$ $\in \{-1,0,+1\}$ are considered. By using inversion symmetry, one can reduce the number of images from 27 to 13. We have no formal proof that the empirical algorithm in Ref.~\cite{militzer16} is rigorous but we compared its predictions with our exact algorithm, presented above, for 10$^9$ randomly sampled cell geometries and atom positions and did not find any deviations.

\bibliography{bibliography}
\bibliographystyle{unsrt}
\end{document}